\documentclass[aps,pra,twocolumn]{revtex4}

\usepackage{graphicx}
\usepackage{setspace}
\usepackage{dcolumn}

\usepackage{color}

\begin{document}

\title{Terahertz-visible two-photon rotational spectroscopy of cold OD$^-$}

\author{Seunghyun Lee}
\author{Daniel Hauser}
\author{Olga Lakhmanskaya}
\author{Steffen Spieler}
\author{Eric S. Endres}
\author{Katharina Geistlinger}
\author{Sunil S. Kumar}
\author{Roland Wester}
\email{roland.wester@uibk.ac.at}
\affiliation{{Institut f{\"u}r Ionenphysik und Angewandte Physik, Universit{\"a}t Innsbruck, Technikerstra{\ss}e 25, 6020 Innsbruck, Austria}}
\date{\today}


\begin{abstract}
We present a method to measure rotational transitions of molecular anions in the terahertz domain by sequential two-photon absorption. Ion excitation by bound-bound terahertz absorption is probed by absorption in the visible on a bound-free transition. The visible frequency is tuned to a state-selective photodetachment transition of the excited anions. This provides a terahertz action spectrum for just few hundred molecular ions. To demonstrate this we measure the two lowest rotational transitions, \textit{J}=1$\leftarrow$0 and \textit{J}=2$\leftarrow$1 of OD$^-$ anions in a cryogenic 22-pole trap. We obtain rotational transition frequencies of 598596.08(19) MHz for \textit{J}=1$\leftarrow$0 and 1196791.57(27) MHz for \textit{J}=2$\leftarrow$1 of OD$^-$, in good agreement with their only previous measurement. This two-photon scheme opens up terahertz rovibrational spectroscopy for a range of molecular anions, in particular for polyatomic and cluster anions.
\end{abstract}
\maketitle


\section{Introduction}

Molecular spectroscopy is currently seeing a lot of development in the terahertz or sub-millimeter wave domain \cite{jepsen2011:lpr,mantsch2010:jms,finneran2015:prl}. This frequency range of 0.3 to 10\,THz started to be explored about thirty years ago \cite{winnewisser1995:vs}, but it has been notoriously difficult to provide sources and detectors for this part of the electromagnetic spectrum. The terahertz range encompasses transitions between rotational states in small molecules such as water or ammonia and between levels of large-amplitude vibrational motion in larger molecules and weakly-bound clusters \cite{moazzen2013:irpc}. A notable example is the vibration-rotation-tunneling transitions in hydrogen-bonded water clusters \cite{keutsch2001:pnas}. The rotational resolution provides access to detailed information on geometrical structures and spin interactions, which allows for precise comparisons with ab initio structure calculations \cite{puzzarini2010:irpc}.

The terahertz domain is not only important for Earth-based spectroscopy and imaging \cite{jepsen2011:lpr}, but also to explore the interstellar medium, in particular the distribution of molecules and ions in cold interstellar and circumstellar molecular clouds \cite{tielens2013:rmp}. With the advent of far-infrared observatories (Herschel, SOFIA, ALMA) this domain has become available for astronomical observations \cite{decin2012:asr}. Provided laboratory data exists, a wealth of information on molecules of interstellar importance is obtained, such as on the molecular ions OH$^+$, OH$_2^+$, OH$_3^+$, ArH$^+$, and H$_2$D$^+$ \cite{gerin2010:aa,barlow2013:sci,bruenken2014:nat}. Laboratory data in the microwave regime have lead to the detection of the interstellar anions C$_4$H$^-$, C$_6$H$^-$, C$_8$H$^-$, CN$^-$, C$_3$N$^-$, and theoretical calculations allowed for the identification of C$_5$N$^-$ \cite{mccarthy2006:apj, cernicharo2007:aa, bruenken2007:apj, kemer2007:apj, thaddeus2008:apj, cernicharo2008:apj, botschwina2008:jpc, agundez2010:aa}. An unsuccessful search has been performed for C$_3$H$^-$ \cite{mcguire2014:apj}.

New technologies are being developed to improve terahertz sources, their resolution, precision, tunability, and intensity, as well as the sensitivity for their detection. Today, the major approaches for continuous-wave generation with narrow linewidth and tunability are quantum cascade lasers \cite{lee2009:book}, frequency multiplication of microwaves \cite{crowe1989:ijim}, and difference frequency generation of near-infrared lasers \cite{deninger2008:rsi,kiessling2013:njp}. The latter approach is also the key to terahertz time-domain spectroscopy \cite{jepsen2011:lpr}. In addition, improvements in terahertz detection sensitivity are being explored, e.\ g.\ cavity enhancement techniques \cite{braakman2011:jap,deprince2013:rsi}, in order to provide new means for the spectroscopy of transient species or for trace gas detection. In the microwave regime, an important advancement in detection signal-to-noise has been achieved with the chirped-pulse technique \cite{dian2008:sci,perez2012:sci}, which is being extended towards the terahertz regime \cite{park2011:jcp}.

Terahertz spectroscopy for molecular ions remains to be a major challenge, because number densities in experiments are orders of magnitude lower than for neutral species. In plasma discharges direct terahertz absorption spectroscopy has been achieved for a selection of small molecular ions: Using diode lasers high lying rotational states of ArH$^+$, NeH$^+$, HeH$^+$, OH$^+$, H$_2$O$^+$, and OH$^-$ have been studied many years ago \cite{liu1987:jcp}. During the following years, rotational states in the frequency regime below 2\,THz have been resolved for several small molecular cations, including SH$^+$ \cite{hovde1987:jcp}, ArH$_3^+$ \cite{bogey1988:jcp}, the hydronium ion H$_3$O$^+$ \cite{verhoeve1989:cpl}, HOCO$^+$ \cite{bogey1988:jms} and HCNH$^+$ \cite{araki1998:apj} and their isotopomers, and C$_2$H$_3^+$ \cite{bogey1992:apj}. The direct spectroscopy of molecular anions in plasma discharges has also received some attention with the studies of SH$^-$ and SD$^-$ \cite{civis1998:jcp} and OH$^-$ and OD$^-$ \cite{cazzoli2005:jcp,cazzoli2006:apj,matsushima2006:jms,yonezu2009:jms}.

The low sensitivity and the difficulty to clearly assign transitions to specific molecular species in these experiments, have stimulated the development of indirect, action-based spectroscopic techniques. In radiofrequency ion traps spectroscopy of mass-selected molecular ions has been developed using state-resolved laser-induced reactions \cite{schlemmer1999:ijm, mikosch2004:jcp, germann2014:natp, asvany2015:sci}, photofragmentation \cite{staanum2010:natp,schneider2010:natp,otto2013:pccp} and - more recently - laser-induced inhibition of cluster growth \cite{chakrabarty2013:jpc,asvany2014:apb} as probing schemes, all of which gain their sensitivity from the detection of the created or remaining number of ions. Up to now this has made it possible to study pure rotational transitions in the terahertz for H$_2$D$^+$, HD$^+$, and OH$^-$ \cite{asvany2008:prl, shen2012:pra, bruenken2014:nat, jusko2014:prl}.

Here we present a method to measure rotational transitions of molecular anions in the terahertz domain by sequential two-photon
absorption. Terahertz absorption on bound-bound transitions is followed by visible absorption on a bound-free transition that leads
to photodetachment of the excess electron. The method is suitable for a large range of negatively charged molecular ions. We demonstrate it by measuring the two lowest rotational transitions in an ensemble of cold, trapped OD$^-$ anions near 0.6 and 1.2\,THz with sub-MHz accuracy in good agreement with the only previous measurement of these transitions \cite{cazzoli2005:jcp, cazzoli2006:apj}. The method is based on quantum state-selective near-threshold photodetachment spectroscopy \cite{otto2013:pccp}. Absorption of a terahertz photon increases the population in the respective upper rotational state, which is detected through photodetachment of the upper state by a second, visible photon. The present approach allows for the study of terahertz transitions across a large frequency range and for a range of negative ions.


\section{Experiment}

\begin{figure}[t!]
\centering
\includegraphics[width=1\columnwidth]{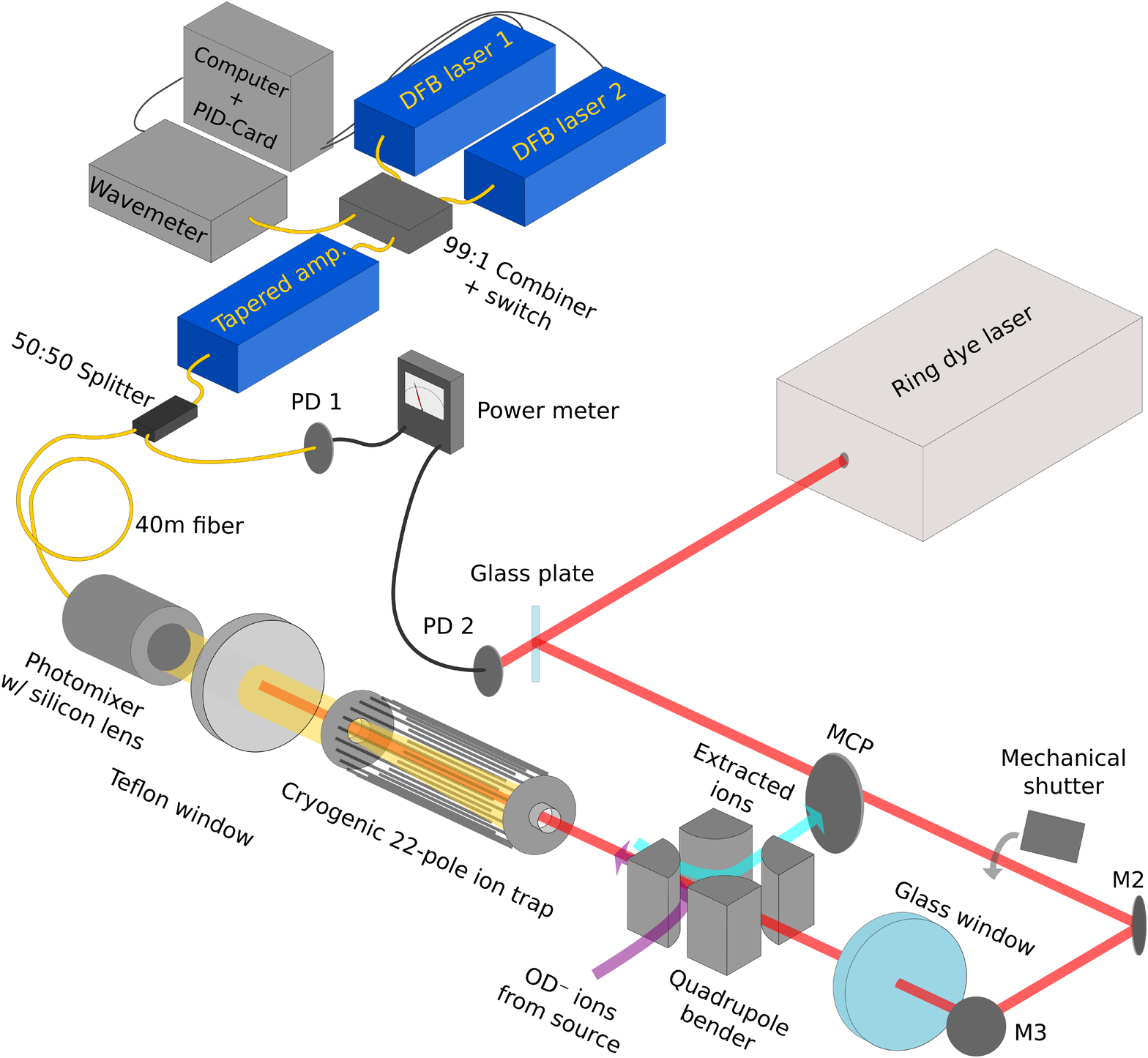}
\caption{Schematic diagram of the major parts of the experimental
  setup. OD$^-$ ions are loaded into a 22-pole ion trap through a
  quadrupole bender and series of ion optics. The ions in the
  trap are exposed to terahertz radiation from an emitter outside the
  vacuum system, driving the rotational transitions
  \textit{J}=1$\leftarrow$0 (\textit{J}=2$\leftarrow$1). A
  ring dye laser for photodetachment, tuned near 680\,nm, is focused into the
  trap to selectively remove population from the excited
  \textit{J}=1(\textit{J}=2) rotational state. Using a mechanical shutter this laser is 
  blocked during ion loading and thermalization. The ion signal that persists after a time interval 
  of photodetachment is measured on a micro-channel plate detector.}
\label{Fig1-Setup}
\end{figure}

The experiment has been performed on OD$^-$ trapped inside a 22-pole radiofrequency ion trap (see Fig.\ \ref{Fig1-Setup}). Tunable terahertz radiation for rotational excitation and a visible laser beam for near-threshold photodetachment are passed into the trap from opposite directions. The photodetachment part of the setup has been described previously \cite{best2011:apj,otto2013:pccp}. Ions are created from a pulsed supersonic expansion of heavy water in a plasma discharge. OD$^-$ ions are mass-selected using a Wiley-McLaren time of flight mass spectrometer. These ions then travel through several ion optics elements and about several hundred of them get loaded into the 22-pole radiofrequency ion trap \cite{gerlich1995:ps,wester2009:jpb}. The trap is mounted on a closed-cycle cryostat at a temperature of $\approx$ 8.5\,K. The trapped ions are cooled by collisions with helium buffer gas that fills the trap at a density of $\approx$ 5$\times 10^{11}$/cm$^3$. The internal temperature acquired by the ions in the cooling process has been determined to be about 20\,K \cite{otto2013:pccp}. It is several Kelvin higher than the trap temperature as known from other studies \cite{mikosch2004:jcp,asvany2008:prl,kreckel2008:jcp,jusko2014:prl}. This is mainly attributed to radiofrequency heating and small imperfections in the trapping field due to surface patch potentials.

The terahertz radiation that is applied to the trapped ions is generated from the difference frequency of two continuous-wave near-infrared diode lasers. We employ a commercially available photomixer-based cw-terahertz system (TOPTICA Photonics) \cite{deninger2008:rsi, saeedkia2013:book}. The system is based on optical heterodyning of two distributed feedback (DFB) diode lasers, as shown in Fig.\ \ref{Fig1-Setup}. A small fraction of each laser beam is fed to a wavelength meter WSU (HighFinesse) for frequency measurement and feedback control of the respective lasers. The main part of the two laser beams is combined using a fiber combiner and intensity-amplified using a tapered diode laser amplifier. The output of the amplifier, carrying the beat signals, is passed onto a photomixer where the terahertz radiation is generated. This radiation is coupled into free space with a silicon lens. The difference frequency signal can be tuned within several seconds across the entire spectral range by changing the temperature of the DFB laser diodes. To check the frequency calibration of the wavelength meter we have performed rotational spectroscopy on neutral CO.

\begin{figure}[t!]
\centering
\includegraphics[width=1\columnwidth]{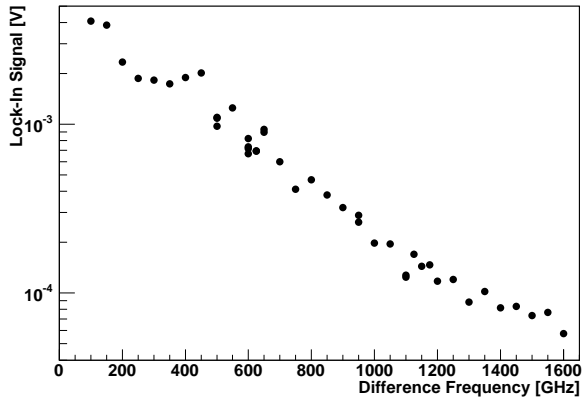}
\caption{Relative intensity of the terahertz radiation source, measured as the lock-in signal of the Golay cell detector, as a function of frequency. The relative intensity is obtained from the lock-in output of the Golay cell. The frequency is taken as the difference frequency of the two infrared lasers measured with the wavemeter.}
\label{Fig2-Power}
\end{figure}

The output power of the terahertz source is estimated with a Golay cell (Tydex) bolometric detector to a range between a few hundred down to a few nanowatts. The lock-in signal from the Golay cell is shown in Fig.\ \ref{Fig2-Power} as a function of the detuning of the two DFB lasers. Terahertz radiation is provided over a wide and continuous tuning range from about 0.1 to 1.6\,THz. The output power decreases roughly as frequency to the third power. At 600\,GHz the absolute terahertz power is estimated to be about 250\,nW at the output of the photomixer, derived by calibrating the Golay cell with near-infrared radiation. The spectral linewidth of the terahertz radiation is derived by measuring the beating signal of the two DFB lasers on a fast photodiode at a difference frequency of only a few GHz. In this way a linewidth of 4\,MHz is determined. This value is the consequence of the finite linewidths of the two lasers. It is in our case given by the linewidth of one of the two DFB laser diodes, which is specified to have a linewidth of a few MHz. The second diode is specified to have a linewidth well below 1\,MHz.

The terahertz emitter is installed along the axial direction of the ion trap close to one of the endcap electrodes, as shown in Fig.\ \ref{Fig1-Setup}. The terahertz radiation is passed into the vacuum system through a teflon window. Inside the ion trap the OD$^-$ ions are exposed to the radiation, which is scanned in frequency around the rotational transition \textit{J}=1$\leftarrow$0 (\textit{J}=2$\leftarrow$1). Any population change in the upper rotational state \textit{J}=1 (\textit{J}=2) is probed by state selective photodetachment of OD$^-$ from that state to neutral OD. This is performed with a tunable cw ring dye laser near 680\,nm (Matisse, Sirah). At a photon energy near 14715\,cm$^{-1}$ (14699 cm$^{-1}$), the laser probes ions that populate the $J=1$ ($J=2$) state. Absorption of the visible photon is determined by measuring the remaining ion signal after extraction from the ion trap on a micro-channel plate detector.


\section{Results}

\begin{figure}[t!]
\centering
\includegraphics[width=1\columnwidth]{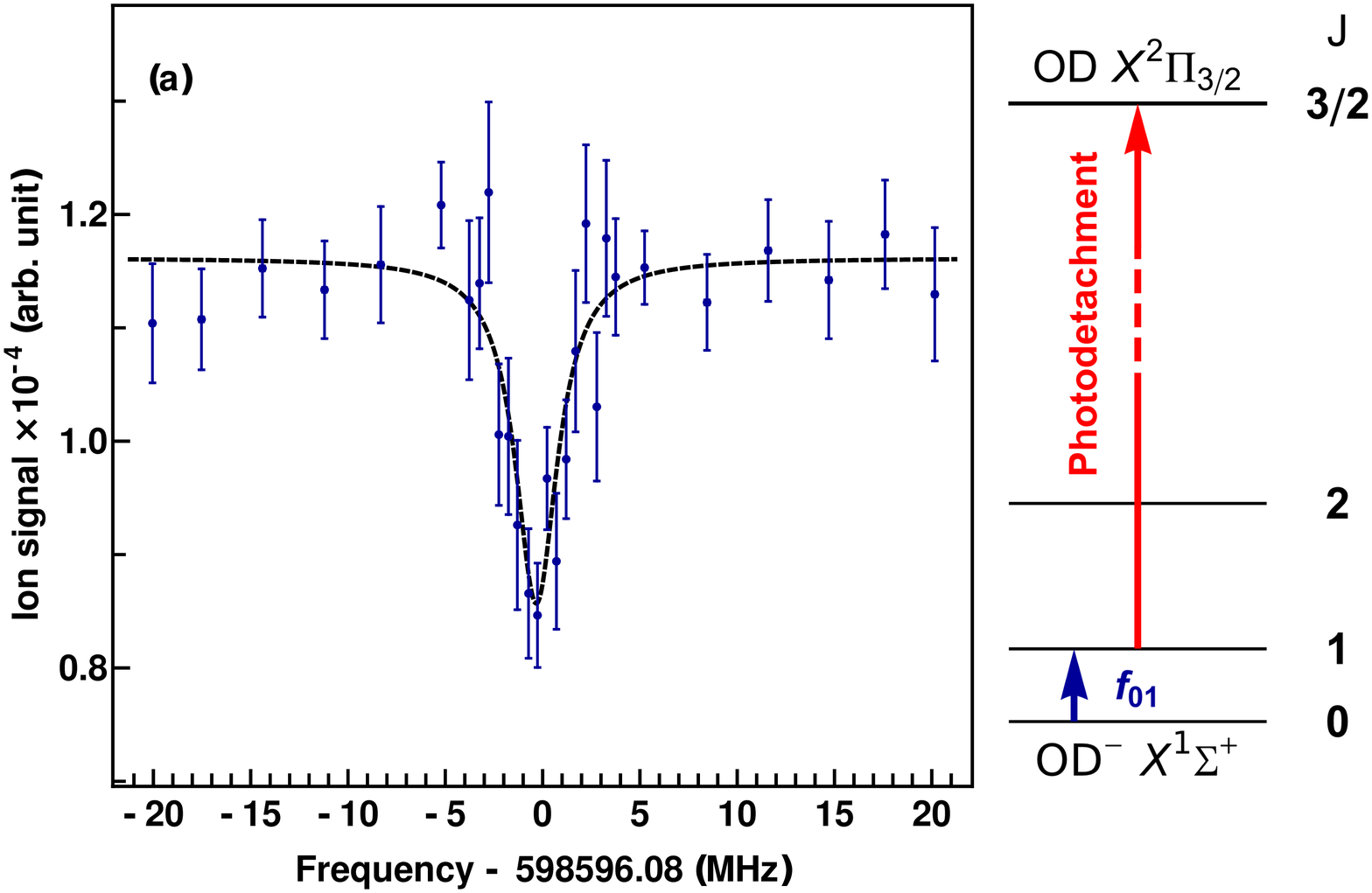}

\vspace{5mm}

\includegraphics[width=1\columnwidth]{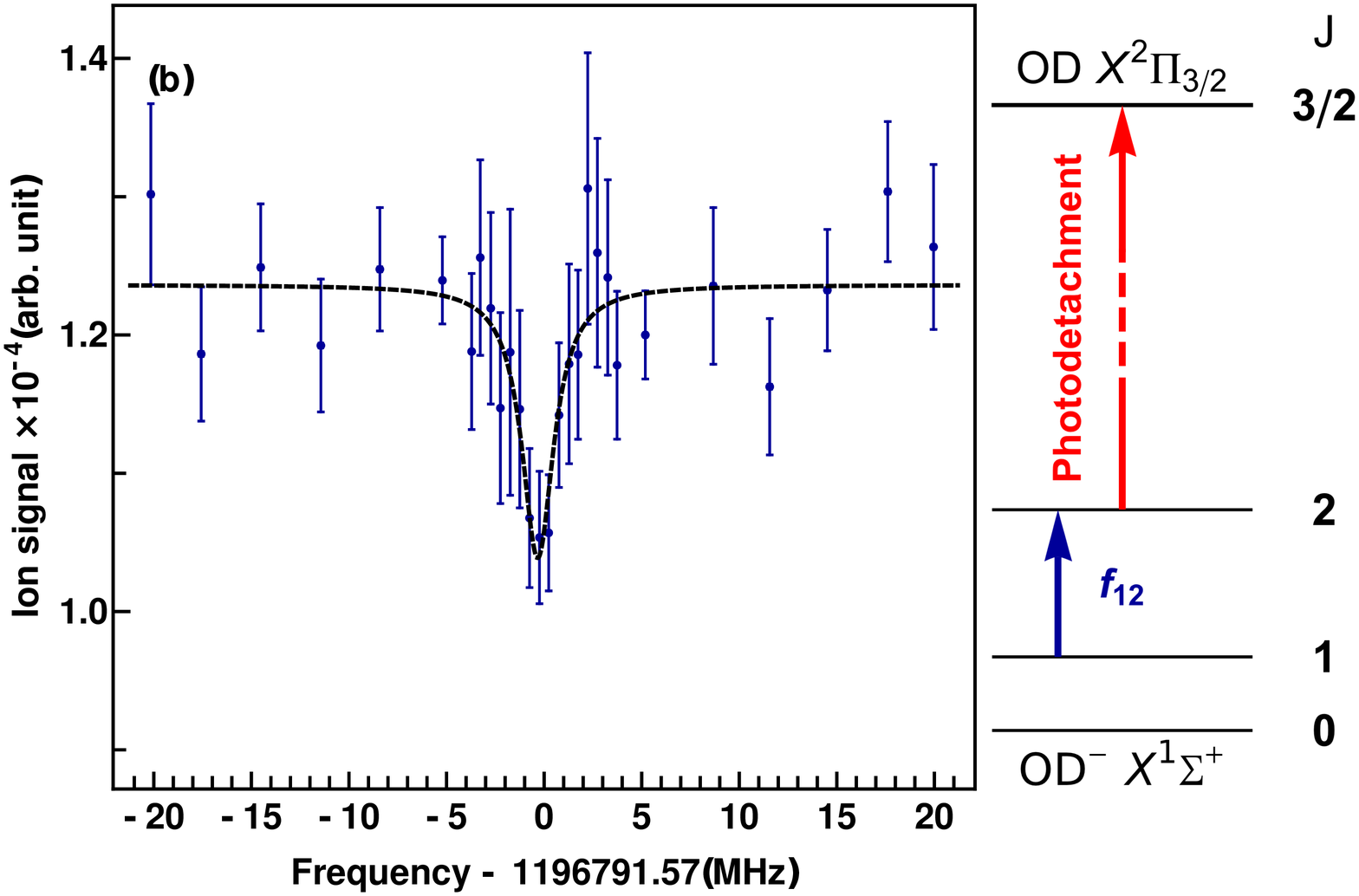}
\caption{Plots of ion signal of OD$^-$ anion after the photodetachment from
  \textit{J}=1 in (a) and \textit{J}=2 in (b) as a function of THz
  radiation frequency. On resonance, ions are pumped into the next higher rotational state
  (\textit{J}=1$\leftarrow$0 in (a) and \textit{J}=2$\leftarrow$1 in
  (b)). The increase in population is probed using state
  selective photodetachment in J=1 in (a) and J=2 in (b). The energy
  manifold (inset) illustrates the THz excitation from \textit{J}=0 to
  1 (\textit{J}=2$\leftarrow$1), marked as a blue arrow and
  photodetachment from \textit{J}=1 (\textit{J}=2) as a red arrow. The
  measured data is fitted using a Lorentzian profile. From the fit,
  the central frequency is obtained as $f_{01}=$ 598596.077(188) MHz
  for \textit{J}=1$\leftarrow$0 and $f_{12}=$ 1196791.573(266)MHz for
  \textit{J}=2$\leftarrow$1.}
\label{Fig3-Spectra}
\end{figure}

Fig.\ \ref{Fig3-Spectra} shows the measured terahertz action spectra. The signals of the remaining OD$^-$ ions in the trap, after photodetachment from \textit{J}=1 and \textit{J}=2, respectively, are plotted in panels (a) and (b) of Fig.\ \ref{Fig3-Spectra} as a function of the terahertz frequency. After a thermalization time of 4\,s during which the photodetachment laser is blocked with the mechanical shutter, the ions have been exposed to this laser for a fixed time of 0.46\,s and 2.3\,s, respectively. These times where chosen, because they correspond to the respective $1/e$ lifetimes of the ions due to photodetachment and yield an optimized signal-to-noise ratio for the terahertz depletion signals. The terahertz radiation is applied continuously. The data points are obtained from the average of 40 spectral scans. The plotted error bars are derived from the standard deviations of the individual measurements. The insets in Fig.\ \ref{Fig3-Spectra} also show the relevant rotational states of OD$^-$, the bound-bound rotational transitions at frequencies $f_{01}$ and $f_{12}$, and the bound-free photodetachment transition to the electron continuum.

The OD$^{-}$ absorption spectra shown in Fig.\ \ref{Fig3-Spectra} have been fitted to a Lorentzian profile, since the linewidth of the terahertz radiation source is dominant over the Doppler width of the ion ensemble, expected to be less than 1\,MHz (FWHM) for both transitions. From the fits, the central frequencies and the statistical precisions for the \textit{J}=1$\leftarrow$0 and \textit{J}=2$\leftarrow$1 transitions are determined to be $f_{01}=598596.08(19)$\,MHz and $f_{12}=1196791.57(27)$\,MHz, respectively. The fitted linewidths amount to 2.6 and 2.0\,MHz (FWHM) for the two transitions. This is slightly smaller than the width determined on a photodiode at a low difference frequency, as presented above. This is attributed to the fact that the linewidth of the spectrally broader laser diode changes with temperature or frequency, respectively, and reaches lower values at higher difference frequencies.


\section{Discussion}

The two studied rotational transitions have previously been observed once in experiments by Cazzoli and Puzzarini, which were performed with a plasma discharge tube \cite{cazzoli2005:jcp, cazzoli2006:apj}. An earlier experiment, which employed rotationally resolved infrared spectroscopy, obtained the transition frequencies with lower resolution \cite{rehfuss1986:jcp}. The comparison of our values with the previous results is shown in Table \ref{table:rot_freq}. The frequencies obtained in this work agree within two standard deviations for the \textit{J}=1$\leftarrow$0 frequency transition and within one standard deviation for the \textit{J}=2$\leftarrow$1 frequency transition. From the two rotational frequencies of OD$^-$ we have calculated the vibration-rotation term values $B_{v=0}$ and $D_{v=0}$ in the ground vibrational level. The calculated values are compared with previous work and listed in Table \ref{table:B0_D0}. Again, our data shows good agreement with the previous results \cite{cazzoli2005:jcp, cazzoli2006:apj,rehfuss1986:jcp,yonezu2009:jms}.

\begin{table*}[bt!]
\caption{Comparison of frequencies (in MHz) of the rotational
  transitions for OD$^{-}$ \textit{J}=1$\leftarrow$0 and
  \textit{J}=2$\leftarrow$1 measured in this work and in previous
  studies:}
\begin{tabular}{c c c c r c c r c c r c }
\hline\hline\\[-1.5ex]
J'$\leftarrow$ J & & & &This work & & & Cazzoli \& Puzzarini\footnote{From Refs.\ ~\cite{cazzoli2005:jcp,cazzoli2006:apj}} & & & Rehfuss \textit{et al.}\footnote{From Ref.\ ~\cite{rehfuss1986:jcp}}&\\
\hline\\[-1.5ex] 
1$\leftarrow$0 & & & & 598596.08(19) & & & 598596.419(63) & & & 598594.4(12.8)& \\
2$\leftarrow$1 & & & &1196791.57(27) & & & 1196791.739(48) & & & 1196789.6(26.6)& \\
\hline\hline 
\end{tabular}
\label{table:rot_freq} 
\end{table*}

\begin{table*}[bt!]
\caption{Comparison of rotational constants (in cm$^{-1}$) of OD$^{-}$
  measured in this work and in previous studies:}
\begin{tabular}{c c c c r c c r c c r c c r c }
\hline\hline\\[-1.5ex] 
 & & & &This work & & & Cazzoli \& Puzzarini \footnote{From Refs. ~\cite{cazzoli2005:jcp,cazzoli2006:apj}} & & & Rehfuss \textit{et al.}\footnote{From Ref.  ~\cite{rehfuss1986:jcp}}& & & Yonezu \textit{et al.}\footnote{From Ref.\ ~\cite{yonezu2009:jms}, combined with data from \cite{cazzoli2005:jcp,cazzoli2006:apj}} & \\
\hline \\[-1.5ex]
$B_0$ & & & & 9.9846214(42) & & & 9.9846286(13) & & & 9.98459(7)& & & 9.9846284(15) &  \\ 
$D_0$ & & & & 5.5675(64)$\times10^{-4}$ & & & 5.5747(11)$\times10^{-4}$& & & 5.549(5)$\times10^{-4}$& & & 5.57571(60)$\times10^{-4}$ &\\
\hline\hline  
\end{tabular}
\label{table:B0_D0} 
\end{table*}

The precision of our transition frequencies is currently still a factor of three to six below that in the study of Cazzoli and Puzzarini \cite{cazzoli2005:jcp, cazzoli2006:apj}. Improvements of our method to reach their precision will be accomplished by improving the signal-to-noise ratio in the experiment. To achieve this a more stable transfer of the ions from the source to the ion trap and a faster data acquisition cycle are currently being implemented. Note, however, that already with the present precision the key benefit of our approach is that a small sample of a few ions, which can be mass-selected prior to the experiment, is sufficient and that reactive collisions do not play a role in the cryogenic helium buffer gas environment.

From the measured depletion signals the intensity of the terahertz radiation in the trap can be estimated. For this purpose the rate equation system for the two lowest rotational states $J=0$ and $1$, coupled by the terahertz radiation, the photodetachment losses of the upper state and the known inelastic collisions \cite{hauser2015:natp} has been solved. For the observed depletion of about 25\% (see Fig.\ \ref{Fig3-Spectra} (a)) a terahertz excitation rate of about 5-10\,/s has been extracted for the fundamental transition. With the dipole matrix element for this transition \cite{leroy:misc}, this rate implies a terahertz power irradiating the ion cloud of about 20-40\,nW near 600\,GHz. This power is more than a factor of ten weaker than the estimated power delivered by the terahertz photomixer, predominantly because of the imperfect transmission of the radiation, with a wavelength of 0.5\,mm, into the ion trap through its 6.6\,mm inner diameter endcap electrodes. These losses will be reduced with an optimized ion trap configuration that is based on a larger separation between the rf rods to allow passing the terahertz radiation radially into the ion trap.

State-selective photodetachment spectroscopy will be applicable to a range of molecular anions. An interesting system to explore is the cluster H$_3$O$_2^-$, the smallest deprotonated water cluster. Its structure is characterized by a central proton, which is delocalized between two classical minimum energy structures in the Born-Oppenheimer potential energy surface. The photodetachment cross section of this system near threshold shows a strong dependence on the trap temperature \cite{otto2013:pccp}, which is indicative for its sensitivity to the population of the rotational and low-lying vibrational states. Rotationally-resolved spectra will provide new and detailed information on the non-classical ground state of this cluster anion.

Spectroscopy in this frequency range is also needed to search for molecular line emission from interstellar molecular clouds and to help understand the chemical networks that lead to the formation of these molecules. Currently, the cation and neutral states of hydroxyl have been observed \cite{gerin2010:aa,wampfler2010:aa}, but not the OH$^-$ and OD$^-$ anions. Nitrogen hydrides, the NH$^+$ and NH$_2^-$ ions, have been searched for and a tentative assignment of the NH$_2^-$ anion has been made at an offset of 141\,MHz, roughly within the estimated accuracy for the transition frequency \cite{persson2014:aa}. Clearly, direct rotational spectroscopy is needed to improve this accuracy and clarify the tentative assignment.

In general, sufficient conditions for the application of our approach to rotational spectroscopy are that photodetachment occurs by s-wave electron emission, which guarantees a high cross section near threshold, and that the Franck-Condon factor for the transition to the vibrational ground state of the neutral is of sufficient strength. Many molecular anions, including polyatomic anions, fulfill these conditions. The frequency range of the method is limited at the low-frequency end by a sufficiently large population difference, which requires a sufficiently low temperature. With an ion temperature of 4\,K, transition frequencies down to below 200\,GHz become accessible. The available terahertz power limits the detection sensitivity at high frequencies to about 1.5\,THz.


\section{Conclusion}

We have presented a method for terahertz action spectroscopy of negatively charged molecules by two-photon detachment action spectroscopy. Using a widely tunable terahertz source, we have measured the \textit{J}=1$\leftarrow$0 and \textit{J}=2$\leftarrow$1 rotational transitions of OD$^{-}$ anions in a 22-pole ion trap using a two photon bound-bound, bound-free spectroscopy scheme. The sensitive detection of terahertz absorption by a small ion ensemble is provided by state-selective near threshold photodetachment. Our experimental results demonstrate a high spectral resolution, which compares well with a previous plasma discharge experiment \cite{cazzoli2005:jcp,cazzoli2006:apj}, and affords a wide tunability of the radiation source over more than one terahertz. Further improvements of the accuracy and the sensitivity of the method are possible. With this, new spectroscopic studies of diatomic and polyatomic molecular anions as well as molecular cluster anions, such as NH$_2^-$ and H$_3$O$_2^-$, become feasible. Such high resolution spectra can provide precision tests of quantum mechanical structure calculations, in particular when tunneling splittings create non-classical ground states. New laboratory data on spectral signatures in the terahertz range may also aid the detection of further molecular ions in the interstellar medium.

\begin{acknowledgments}
This work has been supported by the European Research Council under ERC grant agreement No. 279898. E.S.E. acknowledges support from the Fond National de la Recherche Luxembourg (grant no. 6019121).
\end{acknowledgments}



\begin{thebibliography}{64}
\expandafter\ifx\csname natexlab\endcsname\relax\def\natexlab#1{#1}\fi
\expandafter\ifx\csname bibnamefont\endcsname\relax
  \def\bibnamefont#1{#1}\fi
\expandafter\ifx\csname bibfnamefont\endcsname\relax
  \def\bibfnamefont#1{#1}\fi
\expandafter\ifx\csname citenamefont\endcsname\relax
  \def\citenamefont#1{#1}\fi
\expandafter\ifx\csname url\endcsname\relax
  \def\url#1{\texttt{#1}}\fi
\expandafter\ifx\csname urlprefix\endcsname\relax\def\urlprefix{URL }\fi
\providecommand{\bibinfo}[2]{#2}
\providecommand{\eprint}[2][]{\url{#2}}

\bibitem[{\citenamefont{Jepsen et~al.}(2011)\citenamefont{Jepsen, Cooke, and
  Koch}}]{jepsen2011:lpr}
\bibinfo{author}{\bibfnamefont{P.~U.} \bibnamefont{Jepsen}},
  \bibinfo{author}{\bibfnamefont{D.~G.} \bibnamefont{Cooke}}, \bibnamefont{and}
  \bibinfo{author}{\bibfnamefont{M.}~\bibnamefont{Koch}},
  \bibinfo{journal}{Las. Phot. Rev.} \textbf{\bibinfo{volume}{5}},
  \bibinfo{pages}{124} (\bibinfo{year}{2011}).

\bibitem[{\citenamefont{Mantsch and Naumann}(2010)}]{mantsch2010:jms}
\bibinfo{author}{\bibfnamefont{H.~H.} \bibnamefont{Mantsch}} \bibnamefont{and}
  \bibinfo{author}{\bibfnamefont{D.}~\bibnamefont{Naumann}},
  \bibinfo{journal}{J. Mol. Struct.} \textbf{\bibinfo{volume}{964}},
  \bibinfo{pages}{1} (\bibinfo{year}{2010}).

\bibitem[{\citenamefont{Finneran et~al.}(2015)\citenamefont{Finneran, Good,
  Holland, Carroll, Allodi, and Blake}}]{finneran2015:prl}
\bibinfo{author}{\bibfnamefont{I.~A.} \bibnamefont{Finneran}},
  \bibinfo{author}{\bibfnamefont{J.~T.} \bibnamefont{Good}},
  \bibinfo{author}{\bibfnamefont{D.~B.} \bibnamefont{Holland}},
  \bibinfo{author}{\bibfnamefont{P.~B.} \bibnamefont{Carroll}},
  \bibinfo{author}{\bibfnamefont{M.~A.} \bibnamefont{Allodi}},
  \bibnamefont{and} \bibinfo{author}{\bibfnamefont{G.~A.} \bibnamefont{Blake}},
  \bibinfo{journal}{Phys. Rev. Lett.} \textbf{\bibinfo{volume}{114}},
  \bibinfo{pages}{163902} (\bibinfo{year}{2015}).

\bibitem[{\citenamefont{Winnewisser}(1995)}]{winnewisser1995:vs}
\bibinfo{author}{\bibfnamefont{G.}~\bibnamefont{Winnewisser}},
  \bibinfo{journal}{Vib. Spec.} \textbf{\bibinfo{volume}{8}},
  \bibinfo{pages}{241} (\bibinfo{year}{1995}).

\bibitem[{\citenamefont{Moazzen-Ahmadi and McKellar}(2013)}]{moazzen2013:irpc}
\bibinfo{author}{\bibfnamefont{N.}~\bibnamefont{Moazzen-Ahmadi}}
  \bibnamefont{and} \bibinfo{author}{\bibfnamefont{A.}~\bibnamefont{McKellar}},
  \bibinfo{journal}{Int. Rev. Phys. Chem.} \textbf{\bibinfo{volume}{32}},
  \bibinfo{pages}{611} (\bibinfo{year}{2013}).

\bibitem[{\citenamefont{Keutsch and Saykally}(2001)}]{keutsch2001:pnas}
\bibinfo{author}{\bibfnamefont{F.~N.} \bibnamefont{Keutsch}} \bibnamefont{and}
  \bibinfo{author}{\bibfnamefont{R.~J.} \bibnamefont{Saykally}},
  \bibinfo{journal}{Proc. Nat. Ac. Sci.} \textbf{\bibinfo{volume}{98}},
  \bibinfo{pages}{10533} (\bibinfo{year}{2001}).

\bibitem[{\citenamefont{Puzzarini et~al.}(2010)\citenamefont{Puzzarini,
  Stanton, and Gauss}}]{puzzarini2010:irpc}
\bibinfo{author}{\bibfnamefont{C.}~\bibnamefont{Puzzarini}},
  \bibinfo{author}{\bibfnamefont{J.~F.} \bibnamefont{Stanton}},
  \bibnamefont{and} \bibinfo{author}{\bibfnamefont{J.}~\bibnamefont{Gauss}},
  \bibinfo{journal}{Int. Rev. Phys. Chem.} \textbf{\bibinfo{volume}{29}},
  \bibinfo{pages}{273} (\bibinfo{year}{2010}).

\bibitem[{\citenamefont{Tielens}(2013)}]{tielens2013:rmp}
\bibinfo{author}{\bibfnamefont{A.~G. G.~M.} \bibnamefont{Tielens}},
  \bibinfo{journal}{Rev. Mod. Phys.} \textbf{\bibinfo{volume}{85}},
  \bibinfo{pages}{1021} (\bibinfo{year}{2013}).

\bibitem[{\citenamefont{Decin}(2012)}]{decin2012:asr}
\bibinfo{author}{\bibfnamefont{L.}~\bibnamefont{Decin}}, \bibinfo{journal}{Adv.
  Space Res.} \textbf{\bibinfo{volume}{50}}, \bibinfo{pages}{843}
  (\bibinfo{year}{2012}).

\bibitem[{\citenamefont{Gerin et~al.}(2010)\citenamefont{Gerin, De~Luca, Black,
  Goicoechea, Herbst, Neufeld, Falgarone, Godard, Pearson, Lis
  et~al.}}]{gerin2010:aa}
\bibinfo{author}{\bibfnamefont{M.}~\bibnamefont{Gerin}},
  \bibinfo{author}{\bibfnamefont{M.}~\bibnamefont{De~Luca}},
  \bibinfo{author}{\bibfnamefont{J.}~\bibnamefont{Black}},
  \bibinfo{author}{\bibfnamefont{J.~R.} \bibnamefont{Goicoechea}},
  \bibinfo{author}{\bibfnamefont{E.}~\bibnamefont{Herbst}},
  \bibinfo{author}{\bibfnamefont{D.~A.} \bibnamefont{Neufeld}},
  \bibinfo{author}{\bibfnamefont{E.}~\bibnamefont{Falgarone}},
  \bibinfo{author}{\bibfnamefont{B.}~\bibnamefont{Godard}},
  \bibinfo{author}{\bibfnamefont{J.~C.} \bibnamefont{Pearson}},
  \bibinfo{author}{\bibfnamefont{D.~C.} \bibnamefont{Lis}},
  \bibnamefont{et~al.}, \bibinfo{journal}{Astron. Astrophys.}
  \textbf{\bibinfo{volume}{518}}, \bibinfo{pages}{L110} (\bibinfo{year}{2010}).

\bibitem[{\citenamefont{Barlow et~al.}(2013)\citenamefont{Barlow, Swinyard,
  Owen, Cernicharo, Gomez, Ivison, Krause, Lim, Matsuura, Miller
  et~al.}}]{barlow2013:sci}
\bibinfo{author}{\bibfnamefont{M.}~\bibnamefont{Barlow}},
  \bibinfo{author}{\bibfnamefont{B.}~\bibnamefont{Swinyard}},
  \bibinfo{author}{\bibfnamefont{P.}~\bibnamefont{Owen}},
  \bibinfo{author}{\bibfnamefont{J.}~\bibnamefont{Cernicharo}},
  \bibinfo{author}{\bibfnamefont{H.~L.} \bibnamefont{Gomez}},
  \bibinfo{author}{\bibfnamefont{R.}~\bibnamefont{Ivison}},
  \bibinfo{author}{\bibfnamefont{O.}~\bibnamefont{Krause}},
  \bibinfo{author}{\bibfnamefont{T.}~\bibnamefont{Lim}},
  \bibinfo{author}{\bibfnamefont{M.}~\bibnamefont{Matsuura}},
  \bibinfo{author}{\bibfnamefont{S.}~\bibnamefont{Miller}},
  \bibnamefont{et~al.}, \bibinfo{journal}{Science}
  \textbf{\bibinfo{volume}{342}}, \bibinfo{pages}{1343} (\bibinfo{year}{2013}).

\bibitem[{\citenamefont{Br{\"u}nken et~al.}(2014)\citenamefont{Br{\"u}nken,
  Sipil{\"a}, Chambers, Harju, Caselli, Asvany, Honingh, Kami{\'n}ski, Menten,
  Stutzki et~al.}}]{bruenken2014:nat}
\bibinfo{author}{\bibfnamefont{S.}~\bibnamefont{Br{\"u}nken}},
  \bibinfo{author}{\bibfnamefont{O.}~\bibnamefont{Sipil{\"a}}},
  \bibinfo{author}{\bibfnamefont{E.~T.} \bibnamefont{Chambers}},
  \bibinfo{author}{\bibfnamefont{J.}~\bibnamefont{Harju}},
  \bibinfo{author}{\bibfnamefont{P.}~\bibnamefont{Caselli}},
  \bibinfo{author}{\bibfnamefont{O.}~\bibnamefont{Asvany}},
  \bibinfo{author}{\bibfnamefont{C.~E.} \bibnamefont{Honingh}},
  \bibinfo{author}{\bibfnamefont{T.}~\bibnamefont{Kami{\'n}ski}},
  \bibinfo{author}{\bibfnamefont{K.~M.} \bibnamefont{Menten}},
  \bibinfo{author}{\bibfnamefont{J.}~\bibnamefont{Stutzki}},
  \bibnamefont{et~al.}, \bibinfo{journal}{Nature}
  \textbf{\bibinfo{volume}{516}}, \bibinfo{pages}{219} (\bibinfo{year}{2014}).

\bibitem[{\citenamefont{McCarthy et~al.}(2006)\citenamefont{McCarthy, Gottlieb,
  Gupta, and Thaddeus}}]{mccarthy2006:apj}
\bibinfo{author}{\bibfnamefont{M.~C.} \bibnamefont{McCarthy}},
  \bibinfo{author}{\bibfnamefont{C.~A.} \bibnamefont{Gottlieb}},
  \bibinfo{author}{\bibfnamefont{H.}~\bibnamefont{Gupta}}, \bibnamefont{and}
  \bibinfo{author}{\bibfnamefont{P.}~\bibnamefont{Thaddeus}},
  \bibinfo{journal}{Astrophys. J. Lett.} \textbf{\bibinfo{volume}{652}},
  \bibinfo{pages}{L141} (\bibinfo{year}{2006}).

\bibitem[{\citenamefont{Cernicharo et~al.}(2007)\citenamefont{Cernicharo,
  Gu{\'{e}}lin, Ag{\"u}ndez, Kawaguchi, McCarthy, and
  Thaddeus}}]{cernicharo2007:aa}
\bibinfo{author}{\bibfnamefont{J.}~\bibnamefont{Cernicharo}},
  \bibinfo{author}{\bibfnamefont{M.}~\bibnamefont{Gu{\'{e}}lin}},
  \bibinfo{author}{\bibfnamefont{M.}~\bibnamefont{Ag{\"u}ndez}},
  \bibinfo{author}{\bibfnamefont{K.}~\bibnamefont{Kawaguchi}},
  \bibinfo{author}{\bibfnamefont{M.~C.} \bibnamefont{McCarthy}},
  \bibnamefont{and} \bibinfo{author}{\bibfnamefont{P.}~\bibnamefont{Thaddeus}},
  \bibinfo{journal}{Astron. Astrophys.} \textbf{\bibinfo{volume}{467}},
  \bibinfo{pages}{L37} (\bibinfo{year}{2007}).

\bibitem[{\citenamefont{Br\"unken et~al.}(2007)\citenamefont{Br\"unken, Gupta,
  Gottlieb, McCarthy, and Thaddeus}}]{bruenken2007:apj}
\bibinfo{author}{\bibfnamefont{S.}~\bibnamefont{Br\"unken}},
  \bibinfo{author}{\bibfnamefont{H.}~\bibnamefont{Gupta}},
  \bibinfo{author}{\bibfnamefont{C.~A.} \bibnamefont{Gottlieb}},
  \bibinfo{author}{\bibfnamefont{M.~C.} \bibnamefont{McCarthy}},
  \bibnamefont{and} \bibinfo{author}{\bibfnamefont{P.}~\bibnamefont{Thaddeus}},
  \bibinfo{journal}{Astrophys. J. Lett.} \textbf{\bibinfo{volume}{664}},
  \bibinfo{pages}{L43} (\bibinfo{year}{2007}).

\bibitem[{\citenamefont{Remijan et~al.}(2007)\citenamefont{Remijan, Hollis,
  Lovas, Cordiner, Millar, Markwick-Kemper, and Jewel}}]{kemer2007:apj}
\bibinfo{author}{\bibfnamefont{A.~J.} \bibnamefont{Remijan}},
  \bibinfo{author}{\bibfnamefont{J.~M.} \bibnamefont{Hollis}},
  \bibinfo{author}{\bibfnamefont{F.~J.} \bibnamefont{Lovas}},
  \bibinfo{author}{\bibfnamefont{M.~A.} \bibnamefont{Cordiner}},
  \bibinfo{author}{\bibfnamefont{T.~J.} \bibnamefont{Millar}},
  \bibinfo{author}{\bibfnamefont{A.~J.} \bibnamefont{Markwick-Kemper}},
  \bibnamefont{and} \bibinfo{author}{\bibfnamefont{P.~R.} \bibnamefont{Jewel}},
  \bibinfo{journal}{Astrophys. J.} \textbf{\bibinfo{volume}{664}},
  \bibinfo{pages}{L47} (\bibinfo{year}{2007}).

\bibitem[{\citenamefont{Thaddeus et~al.}(2008)\citenamefont{Thaddeus, Gottlieb,
  Gupta, Br\"{u}nken, and McCarthy}}]{thaddeus2008:apj}
\bibinfo{author}{\bibfnamefont{P.}~\bibnamefont{Thaddeus}},
  \bibinfo{author}{\bibfnamefont{C.~A.} \bibnamefont{Gottlieb}},
  \bibinfo{author}{\bibfnamefont{H.}~\bibnamefont{Gupta}},
  \bibinfo{author}{\bibfnamefont{S.}~\bibnamefont{Br\"{u}nken}},
  \bibnamefont{and} \bibinfo{author}{\bibfnamefont{M.}~\bibnamefont{McCarthy}},
  \bibinfo{journal}{Astrophys. J.} \textbf{\bibinfo{volume}{677}},
  \bibinfo{pages}{1132} (\bibinfo{year}{2008}).

\bibitem[{\citenamefont{Cernicharo et~al.}(2008)\citenamefont{Cernicharo,
  Gu{\'e}lin, Ag{\'u}ndez, McCarthy, and Thaddeus}}]{cernicharo2008:apj}
\bibinfo{author}{\bibfnamefont{J.}~\bibnamefont{Cernicharo}},
  \bibinfo{author}{\bibfnamefont{M.}~\bibnamefont{Gu{\'e}lin}},
  \bibinfo{author}{\bibfnamefont{M.}~\bibnamefont{Ag{\'u}ndez}},
  \bibinfo{author}{\bibfnamefont{M.}~\bibnamefont{McCarthy}}, \bibnamefont{and}
  \bibinfo{author}{\bibfnamefont{P.}~\bibnamefont{Thaddeus}},
  \bibinfo{journal}{Astrophys. J. Lett.} \textbf{\bibinfo{volume}{688}},
  \bibinfo{pages}{L83} (\bibinfo{year}{2008}).

\bibitem[{\citenamefont{Botschwina and Oswald}(2008)}]{botschwina2008:jpc}
\bibinfo{author}{\bibfnamefont{P.}~\bibnamefont{Botschwina}} \bibnamefont{and}
  \bibinfo{author}{\bibfnamefont{R.}~\bibnamefont{Oswald}},
  \bibinfo{journal}{J. Phys. Chem. A} \textbf{\bibinfo{volume}{129}},
  \bibinfo{pages}{044305} (\bibinfo{year}{2008}).

\bibitem[{\citenamefont{Ag{\'u}ndez et~al.}(2010)\citenamefont{Ag{\'u}ndez,
  Cernicharo, Gu{\'e}lin, Kahane, Roueff, K{\l}os, Aoiz, Lique, Marcelino,
  Goicoechea et~al.}}]{agundez2010:aa}
\bibinfo{author}{\bibfnamefont{M.}~\bibnamefont{Ag{\'u}ndez}},
  \bibinfo{author}{\bibfnamefont{J.}~\bibnamefont{Cernicharo}},
  \bibinfo{author}{\bibfnamefont{M.}~\bibnamefont{Gu{\'e}lin}},
  \bibinfo{author}{\bibfnamefont{C.}~\bibnamefont{Kahane}},
  \bibinfo{author}{\bibfnamefont{E.}~\bibnamefont{Roueff}},
  \bibinfo{author}{\bibfnamefont{J.}~\bibnamefont{K{\l}os}},
  \bibinfo{author}{\bibfnamefont{F.~J.} \bibnamefont{Aoiz}},
  \bibinfo{author}{\bibfnamefont{F.}~\bibnamefont{Lique}},
  \bibinfo{author}{\bibfnamefont{N.}~\bibnamefont{Marcelino}},
  \bibinfo{author}{\bibfnamefont{J.~R.} \bibnamefont{Goicoechea}},
  \bibnamefont{et~al.}, \bibinfo{journal}{Astron. Astrophys.}
  \textbf{\bibinfo{volume}{517}}, \bibinfo{pages}{L2} (\bibinfo{year}{2010}).

\bibitem[{\citenamefont{McGuire et~al.}(2014)\citenamefont{McGuire, Carroll,
  Gratier, Guzmán, Pety, Roueff, Gerin, Blake, and Remijan}}]{mcguire2014:apj}
\bibinfo{author}{\bibfnamefont{B.~A.} \bibnamefont{McGuire}},
  \bibinfo{author}{\bibfnamefont{P.~B.} \bibnamefont{Carroll}},
  \bibinfo{author}{\bibfnamefont{P.}~\bibnamefont{Gratier}},
  \bibinfo{author}{\bibfnamefont{V.}~\bibnamefont{Guzmán}},
  \bibinfo{author}{\bibfnamefont{J.}~\bibnamefont{Pety}},
  \bibinfo{author}{\bibfnamefont{E.}~\bibnamefont{Roueff}},
  \bibinfo{author}{\bibfnamefont{M.}~\bibnamefont{Gerin}},
  \bibinfo{author}{\bibfnamefont{G.~A.} \bibnamefont{Blake}}, \bibnamefont{and}
  \bibinfo{author}{\bibfnamefont{A.~J.} \bibnamefont{Remijan}},
  \bibinfo{journal}{Astrophys. J.} \textbf{\bibinfo{volume}{783}},
  \bibinfo{pages}{36} (\bibinfo{year}{2014}).

\bibitem[{\citenamefont{Lee}(2009)}]{lee2009:book}
\bibinfo{author}{\bibfnamefont{Y.~S.} \bibnamefont{Lee}},
  \emph{\bibinfo{title}{Principles of Terahertz Science and Technology}}
  (\bibinfo{publisher}{Springer, NY, USA}, \bibinfo{year}{2009}).

\bibitem[{\citenamefont{Crowe}(1989)}]{crowe1989:ijim}
\bibinfo{author}{\bibfnamefont{T.~W.} \bibnamefont{Crowe}},
  \bibinfo{journal}{Int. J. Infr. Millim. Wav.} \textbf{\bibinfo{volume}{10}},
  \bibinfo{pages}{765} (\bibinfo{year}{1989}).

\bibitem[{\citenamefont{Deninger et~al.}(2008)\citenamefont{Deninger,
  G{\"o}bel, Sch{\"o}nherr, Kinder, Roggenbuck, K{\"o}berle, Lison,
  M{\"u}ller-Wirts, and Meissner}}]{deninger2008:rsi}
\bibinfo{author}{\bibfnamefont{A.~J.} \bibnamefont{Deninger}},
  \bibinfo{author}{\bibfnamefont{T.}~\bibnamefont{G{\"o}bel}},
  \bibinfo{author}{\bibfnamefont{D.}~\bibnamefont{Sch{\"o}nherr}},
  \bibinfo{author}{\bibfnamefont{T.}~\bibnamefont{Kinder}},
  \bibinfo{author}{\bibfnamefont{A.}~\bibnamefont{Roggenbuck}},
  \bibinfo{author}{\bibfnamefont{M.}~\bibnamefont{K{\"o}berle}},
  \bibinfo{author}{\bibfnamefont{F.}~\bibnamefont{Lison}},
  \bibinfo{author}{\bibfnamefont{T.}~\bibnamefont{M{\"u}ller-Wirts}},
  \bibnamefont{and} \bibinfo{author}{\bibfnamefont{P.}~\bibnamefont{Meissner}},
  \bibinfo{journal}{Rev. Sci. Instrum.} \textbf{\bibinfo{volume}{79}},
  \bibinfo{pages}{044702} (\bibinfo{year}{2008}).

\bibitem[{\citenamefont{Kiessling et~al.}(2013)\citenamefont{Kiessling,
  Breunig, Schunemann, Buse, and Vodopyanov}}]{kiessling2013:njp}
\bibinfo{author}{\bibfnamefont{J.}~\bibnamefont{Kiessling}},
  \bibinfo{author}{\bibfnamefont{I.}~\bibnamefont{Breunig}},
  \bibinfo{author}{\bibfnamefont{P.}~\bibnamefont{Schunemann}},
  \bibinfo{author}{\bibfnamefont{K.}~\bibnamefont{Buse}}, \bibnamefont{and}
  \bibinfo{author}{\bibfnamefont{K.}~\bibnamefont{Vodopyanov}},
  \bibinfo{journal}{New J. Phys.} \textbf{\bibinfo{volume}{15}},
  \bibinfo{pages}{105014} (\bibinfo{year}{2013}).

\bibitem[{\citenamefont{Braakman and Blake}(2011)}]{braakman2011:jap}
\bibinfo{author}{\bibfnamefont{R.}~\bibnamefont{Braakman}} \bibnamefont{and}
  \bibinfo{author}{\bibfnamefont{G.}~\bibnamefont{Blake}}, \bibinfo{journal}{J.
  Appl. Phys.} \textbf{\bibinfo{volume}{109}}, \bibinfo{pages}{063102}
  (\bibinfo{year}{2011}).

\bibitem[{\citenamefont{DePrince et~al.}(2013)\citenamefont{DePrince, Rocher,
  Carroll, and Weaver}}]{deprince2013:rsi}
\bibinfo{author}{\bibfnamefont{B.~A.} \bibnamefont{DePrince}},
  \bibinfo{author}{\bibfnamefont{B.~E.} \bibnamefont{Rocher}},
  \bibinfo{author}{\bibfnamefont{A.~M.} \bibnamefont{Carroll}},
  \bibnamefont{and} \bibinfo{author}{\bibfnamefont{S.~L.~W.}
  \bibnamefont{Weaver}}, \bibinfo{journal}{Rev. Sci. Inst.}
  \textbf{\bibinfo{volume}{84}}, \bibinfo{pages}{075107}
  (\bibinfo{year}{2013}).

\bibitem[{\citenamefont{Dian et~al.}(2008)\citenamefont{Dian, Brown, Douglass,
  and Pate}}]{dian2008:sci}
\bibinfo{author}{\bibfnamefont{B.~C.} \bibnamefont{Dian}},
  \bibinfo{author}{\bibfnamefont{G.~G.} \bibnamefont{Brown}},
  \bibinfo{author}{\bibfnamefont{K.~O.} \bibnamefont{Douglass}},
  \bibnamefont{and} \bibinfo{author}{\bibfnamefont{B.~H.} \bibnamefont{Pate}},
  \bibinfo{journal}{Science} \textbf{\bibinfo{volume}{320}},
  \bibinfo{pages}{924} (\bibinfo{year}{2008}).

\bibitem[{\citenamefont{P\'{e}rez et~al.}(2012)\citenamefont{P\'{e}rez, Muckle,
  Zaleski, Seifert, Temelso, Shields, Kisiel, and Pate}}]{perez2012:sci}
\bibinfo{author}{\bibfnamefont{C.}~\bibnamefont{P\'{e}rez}},
  \bibinfo{author}{\bibfnamefont{M.~T.} \bibnamefont{Muckle}},
  \bibinfo{author}{\bibfnamefont{D.~P.} \bibnamefont{Zaleski}},
  \bibinfo{author}{\bibfnamefont{N.~A.} \bibnamefont{Seifert}},
  \bibinfo{author}{\bibfnamefont{D.}~\bibnamefont{Temelso}},
  \bibinfo{author}{\bibfnamefont{G.~C.} \bibnamefont{Shields}},
  \bibinfo{author}{\bibfnamefont{Z.}~\bibnamefont{Kisiel}}, \bibnamefont{and}
  \bibinfo{author}{\bibfnamefont{B.~H.} \bibnamefont{Pate}},
  \bibinfo{journal}{Science} \textbf{\bibinfo{volume}{336}},
  \bibinfo{pages}{897} (\bibinfo{year}{2012}).

\bibitem[{\citenamefont{Park et~al.}(2011)\citenamefont{Park, Steeves,
  Kuyanov-Prozument, Neill, and Field}}]{park2011:jcp}
\bibinfo{author}{\bibfnamefont{G.~B.} \bibnamefont{Park}},
  \bibinfo{author}{\bibfnamefont{A.~H.} \bibnamefont{Steeves}},
  \bibinfo{author}{\bibfnamefont{K.}~\bibnamefont{Kuyanov-Prozument}},
  \bibinfo{author}{\bibfnamefont{J.~L.} \bibnamefont{Neill}}, \bibnamefont{and}
  \bibinfo{author}{\bibfnamefont{R.~W.} \bibnamefont{Field}},
  \bibinfo{journal}{J. Chem. Phys.} \textbf{\bibinfo{volume}{135}},
  \bibinfo{pages}{024202} (\bibinfo{year}{2011}).

\bibitem[{\citenamefont{Liu et~al.}(1987)\citenamefont{Liu, Ho, and
  Oka}}]{liu1987:jcp}
\bibinfo{author}{\bibfnamefont{D.-J.} \bibnamefont{Liu}},
  \bibinfo{author}{\bibfnamefont{W.-C.} \bibnamefont{Ho}}, \bibnamefont{and}
  \bibinfo{author}{\bibfnamefont{T.}~\bibnamefont{Oka}}, \bibinfo{journal}{J.
  Chem. Phys.} \textbf{\bibinfo{volume}{87}}, \bibinfo{pages}{2442}
  (\bibinfo{year}{1987}).

\bibitem[{\citenamefont{Hovde and Saykally}(1987)}]{hovde1987:jcp}
\bibinfo{author}{\bibfnamefont{D.~C.} \bibnamefont{Hovde}} \bibnamefont{and}
  \bibinfo{author}{\bibfnamefont{R.~J.} \bibnamefont{Saykally}},
  \bibinfo{journal}{J. Chem. Phys.} \textbf{\bibinfo{volume}{87}},
  \bibinfo{pages}{4332} (\bibinfo{year}{1987}).

\bibitem[{\citenamefont{Bogey et~al.}(1988{\natexlab{a}})\citenamefont{Bogey,
  Bolvin, Demuynck, Destombes, and Van~Eijck}}]{bogey1988:jcp}
\bibinfo{author}{\bibfnamefont{M.}~\bibnamefont{Bogey}},
  \bibinfo{author}{\bibfnamefont{H.}~\bibnamefont{Bolvin}},
  \bibinfo{author}{\bibfnamefont{C.}~\bibnamefont{Demuynck}},
  \bibinfo{author}{\bibfnamefont{J.~L.} \bibnamefont{Destombes}},
  \bibnamefont{and} \bibinfo{author}{\bibfnamefont{B.~P.}
  \bibnamefont{Van~Eijck}}, \bibinfo{journal}{J. Chem. Phys.}
  \textbf{\bibinfo{volume}{88}}, \bibinfo{pages}{4120}
  (\bibinfo{year}{1988}{\natexlab{a}}).

\bibitem[{\citenamefont{Verhoeve et~al.}(1989)\citenamefont{Verhoeve, Versluis,
  Ter~Meulen, Meerts, and Dymanus}}]{verhoeve1989:cpl}
\bibinfo{author}{\bibfnamefont{P.}~\bibnamefont{Verhoeve}},
  \bibinfo{author}{\bibfnamefont{M.}~\bibnamefont{Versluis}},
  \bibinfo{author}{\bibfnamefont{J.}~\bibnamefont{Ter~Meulen}},
  \bibinfo{author}{\bibfnamefont{W.~L.} \bibnamefont{Meerts}},
  \bibnamefont{and} \bibinfo{author}{\bibfnamefont{A.}~\bibnamefont{Dymanus}},
  \bibinfo{journal}{Chem. Phys. Lett.} \textbf{\bibinfo{volume}{161}},
  \bibinfo{pages}{195} (\bibinfo{year}{1989}).

\bibitem[{\citenamefont{Bogey et~al.}(1988{\natexlab{b}})\citenamefont{Bogey,
  Demuynck, Destombes, and Krupnov}}]{bogey1988:jms}
\bibinfo{author}{\bibfnamefont{M.}~\bibnamefont{Bogey}},
  \bibinfo{author}{\bibfnamefont{C.}~\bibnamefont{Demuynck}},
  \bibinfo{author}{\bibfnamefont{J.~L.} \bibnamefont{Destombes}},
  \bibnamefont{and} \bibinfo{author}{\bibfnamefont{A.}~\bibnamefont{Krupnov}},
  \bibinfo{journal}{J. Mol. Struct.} \textbf{\bibinfo{volume}{190}},
  \bibinfo{pages}{465} (\bibinfo{year}{1988}{\natexlab{b}}).

\bibitem[{\citenamefont{Araki et~al.}(1998)\citenamefont{Araki, Ozeki, and
  Saito}}]{araki1998:apj}
\bibinfo{author}{\bibfnamefont{M.}~\bibnamefont{Araki}},
  \bibinfo{author}{\bibfnamefont{H.}~\bibnamefont{Ozeki}}, \bibnamefont{and}
  \bibinfo{author}{\bibfnamefont{S.}~\bibnamefont{Saito}},
  \bibinfo{journal}{Astrophys. J.} \textbf{\bibinfo{volume}{496}},
  \bibinfo{pages}{L53} (\bibinfo{year}{1998}).

\bibitem[{\citenamefont{Bogey et~al.}(1992)\citenamefont{Bogey, Cordonnier,
  Demuynck, and Destombes}}]{bogey1992:apj}
\bibinfo{author}{\bibfnamefont{M.}~\bibnamefont{Bogey}},
  \bibinfo{author}{\bibfnamefont{M.}~\bibnamefont{Cordonnier}},
  \bibinfo{author}{\bibfnamefont{C.}~\bibnamefont{Demuynck}}, \bibnamefont{and}
  \bibinfo{author}{\bibfnamefont{J.}~\bibnamefont{Destombes}},
  \bibinfo{journal}{Astrophys. J.} \textbf{\bibinfo{volume}{399}},
  \bibinfo{pages}{L103} (\bibinfo{year}{1992}).

\bibitem[{\citenamefont{Civis et~al.}(1998)\citenamefont{Civis, Walters,
  Tretyakov, Bailleux, and Bogey}}]{civis1998:jcp}
\bibinfo{author}{\bibfnamefont{S.}~\bibnamefont{Civis}},
  \bibinfo{author}{\bibfnamefont{A.}~\bibnamefont{Walters}},
  \bibinfo{author}{\bibfnamefont{M.~Y.} \bibnamefont{Tretyakov}},
  \bibinfo{author}{\bibfnamefont{S.}~\bibnamefont{Bailleux}}, \bibnamefont{and}
  \bibinfo{author}{\bibfnamefont{M.}~\bibnamefont{Bogey}}, \bibinfo{journal}{J.
  Chem. Phys.} \textbf{\bibinfo{volume}{108}}, \bibinfo{pages}{8369}
  (\bibinfo{year}{1998}).

\bibitem[{\citenamefont{Cazzoli and Puzzarini}(2005)}]{cazzoli2005:jcp}
\bibinfo{author}{\bibfnamefont{G.}~\bibnamefont{Cazzoli}} \bibnamefont{and}
  \bibinfo{author}{\bibfnamefont{C.}~\bibnamefont{Puzzarini}},
  \bibinfo{journal}{J. Chem. Phys.} \textbf{\bibinfo{volume}{123}},
  \bibinfo{pages}{041101} (\bibinfo{year}{2005}).

\bibitem[{\citenamefont{Cazzoli and Puzzarini}(2006)}]{cazzoli2006:apj}
\bibinfo{author}{\bibfnamefont{G.}~\bibnamefont{Cazzoli}} \bibnamefont{and}
  \bibinfo{author}{\bibfnamefont{C.}~\bibnamefont{Puzzarini}},
  \bibinfo{journal}{Astrophys. J.} \textbf{\bibinfo{volume}{648}},
  \bibinfo{pages}{L79} (\bibinfo{year}{2006}).

\bibitem[{\citenamefont{Matsushima et~al.}(2006)\citenamefont{Matsushima,
  Yonezu, Okabe, Tomaru, and Moriwaki}}]{matsushima2006:jms}
\bibinfo{author}{\bibfnamefont{F.}~\bibnamefont{Matsushima}},
  \bibinfo{author}{\bibfnamefont{T.}~\bibnamefont{Yonezu}},
  \bibinfo{author}{\bibfnamefont{T.}~\bibnamefont{Okabe}},
  \bibinfo{author}{\bibfnamefont{K.}~\bibnamefont{Tomaru}}, \bibnamefont{and}
  \bibinfo{author}{\bibfnamefont{Y.}~\bibnamefont{Moriwaki}},
  \bibinfo{journal}{J. Mol. Spectrosc.} \textbf{\bibinfo{volume}{235}},
  \bibinfo{pages}{261} (\bibinfo{year}{2006}).

\bibitem[{\citenamefont{Yonezu et~al.}(2009)\citenamefont{Yonezu, Matsushima,
  Takahashi, Onmayan, and Moriwaki}}]{yonezu2009:jms}
\bibinfo{author}{\bibfnamefont{T.}~\bibnamefont{Yonezu}},
  \bibinfo{author}{\bibfnamefont{F.}~\bibnamefont{Matsushima}},
  \bibinfo{author}{\bibfnamefont{K.}~\bibnamefont{Takahashi}},
  \bibinfo{author}{\bibfnamefont{J.}~\bibnamefont{Onmayan}}, \bibnamefont{and}
  \bibinfo{author}{\bibfnamefont{Y.}~\bibnamefont{Moriwaki}},
  \bibinfo{journal}{J. Mol. Spectrosc.} \textbf{\bibinfo{volume}{253}},
  \bibinfo{pages}{16} (\bibinfo{year}{2009}).

\bibitem[{\citenamefont{Schlemmer et~al.}(1999)\citenamefont{Schlemmer, Kuhn,
  Lescop, and Gerlich}}]{schlemmer1999:ijm}
\bibinfo{author}{\bibfnamefont{S.}~\bibnamefont{Schlemmer}},
  \bibinfo{author}{\bibfnamefont{T.}~\bibnamefont{Kuhn}},
  \bibinfo{author}{\bibfnamefont{E.}~\bibnamefont{Lescop}}, \bibnamefont{and}
  \bibinfo{author}{\bibfnamefont{D.}~\bibnamefont{Gerlich}},
  \bibinfo{journal}{Int. J. Mass Spectrom.} \textbf{\bibinfo{volume}{185-187}},
  \bibinfo{pages}{589} (\bibinfo{year}{1999}).

\bibitem[{\citenamefont{Mikosch et~al.}(2004)\citenamefont{Mikosch, Kreckel,
  Wester, Plasil, Glosik, Gerlich, Schwalm, and Wolf}}]{mikosch2004:jcp}
\bibinfo{author}{\bibfnamefont{J.}~\bibnamefont{Mikosch}},
  \bibinfo{author}{\bibfnamefont{H.}~\bibnamefont{Kreckel}},
  \bibinfo{author}{\bibfnamefont{R.}~\bibnamefont{Wester}},
  \bibinfo{author}{\bibfnamefont{R.}~\bibnamefont{Plasil}},
  \bibinfo{author}{\bibfnamefont{J.}~\bibnamefont{Glosik}},
  \bibinfo{author}{\bibfnamefont{D.}~\bibnamefont{Gerlich}},
  \bibinfo{author}{\bibfnamefont{D.}~\bibnamefont{Schwalm}}, \bibnamefont{and}
  \bibinfo{author}{\bibfnamefont{A.}~\bibnamefont{Wolf}}, \bibinfo{journal}{J.
  Chem. Phys.} \textbf{\bibinfo{volume}{121}}, \bibinfo{pages}{11030}
  (\bibinfo{year}{2004}).

\bibitem[{\citenamefont{Germann et~al.}(2014)\citenamefont{Germann, Tong, and
  Willitsch}}]{germann2014:natp}
\bibinfo{author}{\bibfnamefont{M.}~\bibnamefont{Germann}},
  \bibinfo{author}{\bibfnamefont{X.}~\bibnamefont{Tong}}, \bibnamefont{and}
  \bibinfo{author}{\bibfnamefont{S.}~\bibnamefont{Willitsch}},
  \bibinfo{journal}{Nature Phys.} \textbf{\bibinfo{volume}{10}},
  \bibinfo{pages}{820} (\bibinfo{year}{2014}).

\bibitem[{\citenamefont{Asvany et~al.}(2015)\citenamefont{Asvany, Yamada,
  Br{\"u}nken, Potapov, and Schlemmer}}]{asvany2015:sci}
\bibinfo{author}{\bibfnamefont{O.}~\bibnamefont{Asvany}},
  \bibinfo{author}{\bibfnamefont{K.~M.~T.} \bibnamefont{Yamada}},
  \bibinfo{author}{\bibfnamefont{S.}~\bibnamefont{Br{\"u}nken}},
  \bibinfo{author}{\bibfnamefont{A.}~\bibnamefont{Potapov}}, \bibnamefont{and}
  \bibinfo{author}{\bibfnamefont{S.}~\bibnamefont{Schlemmer}},
  \bibinfo{journal}{Science} \textbf{\bibinfo{volume}{347}},
  \bibinfo{pages}{1346} (\bibinfo{year}{2015}).

\bibitem[{\citenamefont{Staanum et~al.}(2010)\citenamefont{Staanum,
  H{\o}jbjerre, Skyt, Hansen, and Drewsen}}]{staanum2010:natp}
\bibinfo{author}{\bibfnamefont{P.~F.} \bibnamefont{Staanum}},
  \bibinfo{author}{\bibfnamefont{K.}~\bibnamefont{H{\o}jbjerre}},
  \bibinfo{author}{\bibfnamefont{P.~S.} \bibnamefont{Skyt}},
  \bibinfo{author}{\bibfnamefont{A.~K.} \bibnamefont{Hansen}},
  \bibnamefont{and} \bibinfo{author}{\bibfnamefont{M.}~\bibnamefont{Drewsen}},
  \bibinfo{journal}{Nature Phys.} \textbf{\bibinfo{volume}{6}},
  \bibinfo{pages}{271} (\bibinfo{year}{2010}).

\bibitem[{\citenamefont{Schneider et~al.}(2010)\citenamefont{Schneider, Roth,
  Duncker, Ernsting, and Schiller}}]{schneider2010:natp}
\bibinfo{author}{\bibfnamefont{T.}~\bibnamefont{Schneider}},
  \bibinfo{author}{\bibfnamefont{B.}~\bibnamefont{Roth}},
  \bibinfo{author}{\bibfnamefont{H.}~\bibnamefont{Duncker}},
  \bibinfo{author}{\bibfnamefont{I.}~\bibnamefont{Ernsting}}, \bibnamefont{and}
  \bibinfo{author}{\bibfnamefont{S.}~\bibnamefont{Schiller}},
  \bibinfo{journal}{Nature Phys.} \textbf{\bibinfo{volume}{6}},
  \bibinfo{pages}{275} (\bibinfo{year}{2010}).

\bibitem[{\citenamefont{Otto et~al.}(2013)\citenamefont{Otto, von Zastrow,
  Best, and Wester}}]{otto2013:pccp}
\bibinfo{author}{\bibfnamefont{R.}~\bibnamefont{Otto}},
  \bibinfo{author}{\bibfnamefont{A.}~\bibnamefont{von Zastrow}},
  \bibinfo{author}{\bibfnamefont{T.}~\bibnamefont{Best}}, \bibnamefont{and}
  \bibinfo{author}{\bibfnamefont{R.}~\bibnamefont{Wester}},
  \bibinfo{journal}{Phys. Chem. Chem. Phys.} \textbf{\bibinfo{volume}{15}},
  \bibinfo{pages}{612} (\bibinfo{year}{2013}).

\bibitem[{\citenamefont{Chakrabarty et~al.}(2013)\citenamefont{Chakrabarty,
  Holz, Campbell, Banerjee, Gerlich, and Maier}}]{chakrabarty2013:jpc}
\bibinfo{author}{\bibfnamefont{S.}~\bibnamefont{Chakrabarty}},
  \bibinfo{author}{\bibfnamefont{M.}~\bibnamefont{Holz}},
  \bibinfo{author}{\bibfnamefont{E.~K.} \bibnamefont{Campbell}},
  \bibinfo{author}{\bibfnamefont{A.}~\bibnamefont{Banerjee}},
  \bibinfo{author}{\bibfnamefont{D.}~\bibnamefont{Gerlich}}, \bibnamefont{and}
  \bibinfo{author}{\bibfnamefont{J.~P.} \bibnamefont{Maier}},
  \bibinfo{journal}{J. Phys. Chem. Lett.} \textbf{\bibinfo{volume}{4}},
  \bibinfo{pages}{4051} (\bibinfo{year}{2013}).

\bibitem[{\citenamefont{Asvany et~al.}(2014)\citenamefont{Asvany, B{\"u}nken,
  Kluge, and Schlemmer}}]{asvany2014:apb}
\bibinfo{author}{\bibfnamefont{O.}~\bibnamefont{Asvany}},
  \bibinfo{author}{\bibfnamefont{S.}~\bibnamefont{B{\"u}nken}},
  \bibinfo{author}{\bibfnamefont{L.}~\bibnamefont{Kluge}}, \bibnamefont{and}
  \bibinfo{author}{\bibfnamefont{S.}~\bibnamefont{Schlemmer}},
  \bibinfo{journal}{Appl. Phys. B} \textbf{\bibinfo{volume}{114}},
  \bibinfo{pages}{203} (\bibinfo{year}{2014}).

\bibitem[{\citenamefont{Asvany et~al.}(2008)\citenamefont{Asvany, Ricken,
  M{\"u}ller, Wiedner, Giesen, and Schlemmer}}]{asvany2008:prl}
\bibinfo{author}{\bibfnamefont{O.}~\bibnamefont{Asvany}},
  \bibinfo{author}{\bibfnamefont{O.}~\bibnamefont{Ricken}},
  \bibinfo{author}{\bibfnamefont{H.~S.~P.} \bibnamefont{M{\"u}ller}},
  \bibinfo{author}{\bibfnamefont{M.~C.} \bibnamefont{Wiedner}},
  \bibinfo{author}{\bibfnamefont{T.~F.} \bibnamefont{Giesen}},
  \bibnamefont{and}
  \bibinfo{author}{\bibfnamefont{S.}~\bibnamefont{Schlemmer}},
  \bibinfo{journal}{Phys. Rev. Lett.} \textbf{\bibinfo{volume}{100}},
  \bibinfo{pages}{233004} (\bibinfo{year}{2008}).

\bibitem[{\citenamefont{Shen et~al.}(2012)\citenamefont{Shen, Borodin, Hansen,
  and Schiller}}]{shen2012:pra}
\bibinfo{author}{\bibfnamefont{J.}~\bibnamefont{Shen}},
  \bibinfo{author}{\bibfnamefont{A.}~\bibnamefont{Borodin}},
  \bibinfo{author}{\bibfnamefont{M.}~\bibnamefont{Hansen}}, \bibnamefont{and}
  \bibinfo{author}{\bibfnamefont{S.}~\bibnamefont{Schiller}},
  \bibinfo{journal}{Phys. Rev. A} \textbf{\bibinfo{volume}{85}},
  \bibinfo{pages}{032519} (\bibinfo{year}{2012}).

\bibitem[{\citenamefont{Jusko et~al.}(2014)\citenamefont{Jusko, Asvany,
  Wallerstein, Br{\"u}nken, and Schlemmer}}]{jusko2014:prl}
\bibinfo{author}{\bibfnamefont{P.}~\bibnamefont{Jusko}},
  \bibinfo{author}{\bibfnamefont{O.}~\bibnamefont{Asvany}},
  \bibinfo{author}{\bibfnamefont{A.-C.} \bibnamefont{Wallerstein}},
  \bibinfo{author}{\bibfnamefont{S.}~\bibnamefont{Br{\"u}nken}},
  \bibnamefont{and}
  \bibinfo{author}{\bibfnamefont{S.}~\bibnamefont{Schlemmer}},
  \bibinfo{journal}{Phys. Rev. Lett.} \textbf{\bibinfo{volume}{112}},
  \bibinfo{pages}{253005} (\bibinfo{year}{2014}).

\bibitem[{\citenamefont{Best et~al.}(2011)\citenamefont{Best, Otto, Trippel,
  Hlavenka, von Zastrow, Eisenbach, Jezouin, Wester, Vigren, Hamberg
  et~al.}}]{best2011:apj}
\bibinfo{author}{\bibfnamefont{T.}~\bibnamefont{Best}},
  \bibinfo{author}{\bibfnamefont{R.}~\bibnamefont{Otto}},
  \bibinfo{author}{\bibfnamefont{S.}~\bibnamefont{Trippel}},
  \bibinfo{author}{\bibfnamefont{P.}~\bibnamefont{Hlavenka}},
  \bibinfo{author}{\bibfnamefont{A.}~\bibnamefont{von Zastrow}},
  \bibinfo{author}{\bibfnamefont{S.}~\bibnamefont{Eisenbach}},
  \bibinfo{author}{\bibfnamefont{S.}~\bibnamefont{Jezouin}},
  \bibinfo{author}{\bibfnamefont{R.}~\bibnamefont{Wester}},
  \bibinfo{author}{\bibfnamefont{E.}~\bibnamefont{Vigren}},
  \bibinfo{author}{\bibfnamefont{M.}~\bibnamefont{Hamberg}},
  \bibnamefont{et~al.}, \bibinfo{journal}{Ap. J.}
  \textbf{\bibinfo{volume}{742}}, \bibinfo{pages}{63} (\bibinfo{year}{2011}).

\bibitem[{\citenamefont{Gerlich}(1995)}]{gerlich1995:ps}
\bibinfo{author}{\bibfnamefont{D.}~\bibnamefont{Gerlich}},
  \bibinfo{journal}{Phys. Scr.} \textbf{\bibinfo{volume}{T59}},
  \bibinfo{pages}{256} (\bibinfo{year}{1995}).

\bibitem[{\citenamefont{Wester}(2009)}]{wester2009:jpb}
\bibinfo{author}{\bibfnamefont{R.}~\bibnamefont{Wester}}, \bibinfo{journal}{J.
  Phys. B} \textbf{\bibinfo{volume}{42}}, \bibinfo{pages}{154001}
  (\bibinfo{year}{2009}).

\bibitem[{\citenamefont{Kreckel et~al.}(2008)\citenamefont{Kreckel, Bing,
  Reinhardt, Petrignani, Berg, and Wolf}}]{kreckel2008:jcp}
\bibinfo{author}{\bibfnamefont{H.}~\bibnamefont{Kreckel}},
  \bibinfo{author}{\bibfnamefont{D.}~\bibnamefont{Bing}},
  \bibinfo{author}{\bibfnamefont{S.}~\bibnamefont{Reinhardt}},
  \bibinfo{author}{\bibfnamefont{A.}~\bibnamefont{Petrignani}},
  \bibinfo{author}{\bibfnamefont{M.}~\bibnamefont{Berg}}, \bibnamefont{and}
  \bibinfo{author}{\bibfnamefont{A.}~\bibnamefont{Wolf}}, \bibinfo{journal}{J.
  Chem. Phys.} \textbf{\bibinfo{volume}{129}}, \bibinfo{pages}{164312}
  (\bibinfo{year}{2008}).

\bibitem[{\citenamefont{Saeedkia}(2013)}]{saeedkia2013:book}
\bibinfo{author}{\bibfnamefont{D.}~\bibnamefont{Saeedkia}},
  \emph{\bibinfo{title}{Handbook of terahertz technology for imaging, sensing
  and communications}} (\bibinfo{publisher}{Woodhead Publishing Ltd.,
  Cambridge, UK}, \bibinfo{year}{2013}).

\bibitem[{\citenamefont{Rehfuss et~al.}(1986)\citenamefont{Rehfuss, Crofton,
  and Oka}}]{rehfuss1986:jcp}
\bibinfo{author}{\bibfnamefont{B.~D.} \bibnamefont{Rehfuss}},
  \bibinfo{author}{\bibfnamefont{M.~W.} \bibnamefont{Crofton}},
  \bibnamefont{and} \bibinfo{author}{\bibfnamefont{T.}~\bibnamefont{Oka}},
  \bibinfo{journal}{J. Chem. Phys.} \textbf{\bibinfo{volume}{85}},
  \bibinfo{pages}{1785} (\bibinfo{year}{1986}).

\bibitem[{\citenamefont{Hauser et~al.}(2015)\citenamefont{Hauser, Lee, Carelli,
  Spieler, Lakhmanskaya, Endres, Kumar, Gianturco, and
  Wester}}]{hauser2015:natp}
\bibinfo{author}{\bibfnamefont{D.}~\bibnamefont{Hauser}},
  \bibinfo{author}{\bibfnamefont{S.}~\bibnamefont{Lee}},
  \bibinfo{author}{\bibfnamefont{F.}~\bibnamefont{Carelli}},
  \bibinfo{author}{\bibfnamefont{S.}~\bibnamefont{Spieler}},
  \bibinfo{author}{\bibfnamefont{O.}~\bibnamefont{Lakhmanskaya}},
  \bibinfo{author}{\bibfnamefont{E.~S.} \bibnamefont{Endres}},
  \bibinfo{author}{\bibfnamefont{S.~S.} \bibnamefont{Kumar}},
  \bibinfo{author}{\bibfnamefont{F.}~\bibnamefont{Gianturco}},
  \bibnamefont{and} \bibinfo{author}{\bibfnamefont{R.}~\bibnamefont{Wester}},
  \bibinfo{journal}{Nature Phys.} \textbf{\bibinfo{volume}{11}},
  \bibinfo{pages}{467} (\bibinfo{year}{2015}).

\bibitem[{ler()}]{leroy:misc}
\bibinfo{note}{Einstein A, B-coefficients are computed with LEVEL 7.5 by R. J.
  LeRoy (university of Waterloo) based on the ground state dipole moment
  function provided by G. C. Groenenboom (Radboud University, Nijmegen)}.

\bibitem[{\citenamefont{Wampfler et~al.}(2010)\citenamefont{Wampfler, Herczeg,
  Bruderer, Benz, van Dishoeck, Kristensen, Visser, Doty, Melchior, van Kempen
  et~al.}}]{wampfler2010:aa}
\bibinfo{author}{\bibfnamefont{S.~F.} \bibnamefont{Wampfler}},
  \bibinfo{author}{\bibfnamefont{G.~J.} \bibnamefont{Herczeg}},
  \bibinfo{author}{\bibfnamefont{S.}~\bibnamefont{Bruderer}},
  \bibinfo{author}{\bibfnamefont{A.~O.} \bibnamefont{Benz}},
  \bibinfo{author}{\bibfnamefont{E.~F.} \bibnamefont{van Dishoeck}},
  \bibinfo{author}{\bibfnamefont{L.~E.} \bibnamefont{Kristensen}},
  \bibinfo{author}{\bibfnamefont{R.}~\bibnamefont{Visser}},
  \bibinfo{author}{\bibfnamefont{S.~D.} \bibnamefont{Doty}},
  \bibinfo{author}{\bibfnamefont{M.}~\bibnamefont{Melchior}},
  \bibinfo{author}{\bibfnamefont{T.~A.} \bibnamefont{van Kempen}},
  \bibnamefont{et~al.}, \bibinfo{journal}{Astron. Astrophys.}
  \textbf{\bibinfo{volume}{521}}, \bibinfo{pages}{L36} (\bibinfo{year}{2010}).

\bibitem[{\citenamefont{Persson et~al.}(2014)\citenamefont{Persson, Hajigholi,
  Hassel, Olofsson, Black, Herbst, M{\"u}ller, Cernicharo, Wirstr{\"o}m, Olberg
  et~al.}}]{persson2014:aa}
\bibinfo{author}{\bibfnamefont{C.~M.} \bibnamefont{Persson}},
  \bibinfo{author}{\bibfnamefont{M.}~\bibnamefont{Hajigholi}},
  \bibinfo{author}{\bibfnamefont{G.}~\bibnamefont{Hassel}},
  \bibinfo{author}{\bibfnamefont{A.}~\bibnamefont{Olofsson}},
  \bibinfo{author}{\bibfnamefont{J.~H.} \bibnamefont{Black}},
  \bibinfo{author}{\bibfnamefont{E.}~\bibnamefont{Herbst}},
  \bibinfo{author}{\bibfnamefont{H.}~\bibnamefont{M{\"u}ller}},
  \bibinfo{author}{\bibfnamefont{J.}~\bibnamefont{Cernicharo}},
  \bibinfo{author}{\bibfnamefont{E.}~\bibnamefont{Wirstr{\"o}m}},
  \bibinfo{author}{\bibfnamefont{M.}~\bibnamefont{Olberg}},
  \bibnamefont{et~al.}, \bibinfo{journal}{Astron. Astrophys.}
  \textbf{\bibinfo{volume}{567}}, \bibinfo{pages}{A130} (\bibinfo{year}{2014}).

\end{thebibliography}
\end{document}